# Parallel bi-objective evolutionary algorithms for scalable feature subset selection via migration strategy under Spark

## Yelleti Vivek[1,2], Vadlamani Ravi[1*] and P. Radha Krishna[2]

[1]Analytics Lab, Centre for Artificial Intelligence and Machine Learning
Institute for Development and Research in Banking Technology,
Castle Hills Road #1, Masab Tank, Hyderabad-500076, India
[2]Department of Computer Science and Engineering, National Institute of Technology,
Warangal-506004, India
yvivek@idrbt.ac.in; padmarav@idrbt.ac.in; prkrishna@nitw.ac.in

**Abstract**

Feature subset selection (FSS) for classification is inherently a bi-objective optimization problem, where the task is to obtain a feature subset which yields the maximum possible area under the receiver operator characteristic curve (AUC) with minimum cardinality of the feature subset. In today's world, humungous amount of data is generated in all activities of humans. To mine such voluminous data, which is often high-dimensional, there is a need to develop parallel and scalable frameworks. In the first-of-its-kind study, we propose and develop an iterative MapReduce based framework for bi-objective evolutionary algorithms (EAs) based wrappers under Apache spark with the migration strategy. In order to accomplish this, we parallelized the non-dominated sorting based algorithms namely non dominated sorting algorithm (NSGA-II), and non-dominated sorting particle swarm optimization (NSPSO), also the decomposition based algorithm, namely the multi objective evolutionary algorithm based on decomposition (MOEA/D), and named them P-NSGA-II-IS, P-NSPSO-IS, P-MOEA/D-IS, respectively. We proposed a modified MOEA/D by incorporating the non-dominated sorting principle while paralleizing it. Throughout the study, AUC is computed by logistic regression (LR). We test the effectiveness of the proposed methodology on various datasets. It is noteworthy that the P-NSGA-II turns out to be statistically significant by being in the top 2 positions on most datasets. We also reported the empirical attainment plots, speedup analysis, and mean AUC obtained by the most repeated feature subset and the least cardinal feature subset with the highest AUC, and diversity analysis using hypervolume.

# 1. Introduction

After the advent of Web 2.0 and big data, real-world problems became replete with a vast number of dimensions/features. In such cases, there is a possibility that all features are not highly discriminative, informative and not equally important. This implies that in those datasets, there exist some irrelevant and redundant features. Including this kind of features will increase the model training cost, decreases the performance of the models trained on them and decreases human comprehensibility.. Feature subset selection (FSS) [1,2], which alleviates such problems, is the process of selecting the highly discriminative, informative and relevant feature subset. FSS is an essential step in CRISP-DM [45] framework of building models for data mining such as classification, regression and clustering.

FSS [2] can be accomplished in the following three ways viz., (i) filter, (ii) wrapper, and (iii) embedded approaches. Filter approaches selects the features based on the performance of the statistical measure neverthless of the employed model. The most common measures are viz., Information gain, Gain ratio, Symmetrical uncertainty etc., Typically filters are the least computationally expensive than the other two methods yet shows disadvantage by achieving less accuracy. Wrapper [35] approaches involve selecting features for a machine learning model that is used to build on a given dataset. Typically, for the classification tasks, wrapper approaches evaluate the strength of the feature subsets based on the classifier performance (eg: Naive Bayes, Logisitic regression etc.,). On the other hand, approaches where the feature subset selection is included as a part of model-building phase, are considered as embedded approaches. These approaches, in general, entail less accuracy while being computationally less expensive. Eventhough the wrappers are more complex than filter approaches yet the former approaches outperform the latter [67]. This is due to the fact that the wrappers effectively consider the interaction among the features than the filters. Based on the above discussed advantages, we decided to employ the wrapper approach in the current study.

---
* Corresponding Author

FSS is essentially a combinatorial optimization problem [35]. Considering *n* original features in a dataset, the possible number of feature combinations is $2^n - 1$. As *n* becomes large, FSS becomes a and hence, performing an exhaustive search is practically cumbersome. The classical optimization algorithms, which often suffer from entrapment in the local optimal solutions, are not suitable to apply on the multi-modal problems etc. Evolutionary algorithms (EAs) circumvent these hurdles and are proven to solve many optimization problems.

The major objective of FSS is to obtain the maximum possible AUC with the least cardinal feature subset. Accordingly, the FSS can be formulated as a multi objective optimization problem (MOPs). Here, AUC and cardinality of the corresponding feature subset are considered as two objective functions. Multi objective evolutionary algorithms (MOEAs) are proven to be very effective in solving various combinatorial and continuous MOPs. MOEAs are majorly divided into three different categories: (i) pareto dominance based algorithms: here, the conflict between the algorithms can be in as the Non-dominating sorting (NS) principle. For eg: NSGA-II [32], NSPSO [33], Strength pareto Evolutionary Algorithm (SPEA), SPEA2 etc, (ii) Decomposition based algorithms: here, the MOPs are decomposed into single objective problems (SOPs) and solved independently. Thereafter, they are aggregated using weighted mechanism principles such as weighted approach, tchebycheff approach, For eg: MOEA/D [34], etc. (iii) Indicator based MOEAs: the performance indicators are integrated into the MOEAs. For eg: Indicator based Evolutionary Algorithm (IBEA), Steady State Metric Selection Evolutionary Multi-Objective Optimization Algorithms (SMS-EMOA), etc., In the current study, we considered NSGA-II, NSPSO and MOEA/D because they are well time-tested and proven to be effective in solving various MOPs. Moreover, in solving FSS, several variants of NSGA-II, PSO, and MOEA/D are proposed in a sequential environment and are proven to be effective (see Table 1).

Nowadays, the data is being generated massively in large amounts and often the real world datasets are high dimensional. This increased the need to design scalable solutions to analyze high-dimensional big datasets. These emerging challenges sparked the idea of designing the new tools, algorithms, and methodologies. MapReduce [36] works on the principle of map and reduce, which is a programming paradigm that is quite popular in handling big data. Recently, spark [37], is an open-source, is emerged as an important big data tool in handling massive big data. spark has the following advantages: fast computing due to its in-memory computation capability, lazy evaluation, and immutability. Moreover, spark also integrates with other big data tools such as Hadoop.

As discussed earlier, MOEAs are very effective in solving MOPs. But, as the number of objective functions increases, the computational complexity involved also increases dramatically. This motivated us to develop the MOEA based wrapper to meet the current demand of analyzing high-dimensional big dataset. Many parallel MOEAs have been proposed in recent times to solve various problems as presented in Table 2. Parallel MOEAs are broadly classified into two strategies: master-slave and island models.

(a) Master-slave model: Only one global population is maintained and the model is implemented synchronously or asynchronously. The significant difference between these two is that in synchronous methods, all the sub-populations evolve simultaneously in every generation. In contrast, in the asynchronous methods, all the sub-populations evolve separetly.

(b) Island model: This approach is also called a distributed model as it is usually implemented in the distributed memory. Each island contains a sub-population that evolves independently, and then the migration policy is applied to periodically interact with the neighbouring solutions evolved from the other islands.

Between these two models, in terms of computational complexity and suitability to the distributed environment, the island models are more suitable to the distributed environment like spark, which works independently of the underlying topology. Hence, we have chosen island models for consideration in the current work.

Although several MOEA based wrapper algorithms are proposed to solve FSS, they are all sequential in implementation. Extant methods cannot analyze big data sets efficiently, resulting in poor performance of the model. To meet the current demands, there is a need to develop scalable wrappers for FSS in the context of big data. These motivated us to propose a scalable wrapper for FSS in the bi-objective environment. To the best of our knowledge, only two works are reported developing parallel EA-based wrappers in the Apache spark but developed in the single objective environment [43,44]. Further, to the best of our knowledge, little or no work is reported dealing with parallel MOEA based wrapper under the spark environment.

In the current study, we posed FSS as a bi-objective optimization problem, where AUC and cardinality of the feature subset are considered as two objective functions to be maximized and minimized respectively. This is because the objective of FSS is to select the feature subset with least possible cardinality that achieves maximum AUC. The current study recognized the importance of the two objectives majorly because:

i) In a single objective, focus is on only one objective function and its individual global maximum or near-global optimal solution is obtained, whereas in the bi-objective case, the importance of feature subset cardinality is also accorded due prominence and the algorithms entails a set of non-dominated set of solutions. This set of non-dominated solutions provides a lot of flexibility to the decision maker.

ii) Further, the trade-off between the objective functions i.e., AUC and cardinality in this context can also be analyzed well.

The highlights of the current work are as follows:
- Proposed and designed an island based approach under apache spark, which is suitable to accommodate the EA based wrapper under a bi-objective environment. In this framework, logistic regression (LR) is chosen as the classifier to evaluate one of the fitness functions, namely the AUC, and the cardinality, which is the number of the features in the selected feature subset.
- Designed the island based parallel EA based wrappers with the popular bi-objective EA algorithms such as NSGA-II, NSPSO, and MOEA/D and analyzed their performance.
- Proposed a modified MOEA/D by incorporating the NS principle via migration policy while paralleizing it.
- Proposed and designed the migration strategy, based on the non-dominating sorting (NS) principle, which is suitable for the bi-objective algorithms in this framework.

The remainder of the paper is sorted out as follows: Section 2 presents the literature review. Section 3 presents the background theory related to the employed heuristics. Section 4 presents the proposed parallel methodology designed under the Apache spark. Section 5 describes the datasets. Section 6 discusses the results obtained by various methods. Finally, section 7 concludes the paper.

Table 1: Table showing the MOEAs solving FSS

| Authors | Algorithm | Filter/Wrapper (Classifier) | Objective Functions | Sequential / Parallel |
|---|---|---|---|---|
| Khammasi and Krichen [46] | NSGA-II | Wrapper (logistic regression) | • Maximize accuracy<br>• Minimize cardinality of the selected feature subset | Sequential |
| Pacheco et al. [47] | NSGA-II | Wrapper (fisher) | Minimize misclassification error rates corresponding to each class:<br>• Type-I error rate<br>• Type II error rate | Sequential |
| Das and Das [48] | MOEA/D | Filter | • Maximize relevance score<br>• Minimize redundancy score | Sequential |
| Huang et al. [49] | NSGA-II | Wrapper (Decision Tree C4.5) | • Maximize true positive (TP)<br>• Maximize true negative (TN) | Sequential |
| Amoozegar et al. [50] | PSO based multi-objective method for FS (RSPDOFS) | Wrapper (KNN) | • Minimize error rate<br>• Minimize cardinality of the selected feature subset | Sequential |
| Peimankar et al. [51] | PSO based multi-objective (MOPSO) | Wrapper (ensemble models) | • Minimize error rate<br>• Minimize cardinality of the selected feature subset | Sequential |
| Sohrabi and Tajik [52] | NSGA-II and MOPSO | Wrapper (Multi-layer perceptron – MLP) | • Minimize MLP Error<br>• Minimize cardinality of the selected feature subset | Sequential |
| Zhang et al. [53] | MOPSO | Wrapper (MLP) | • Maximize accuracy<br>• Minimize cardinality of the selected feature subset | Sequential |
| Yong et al. [54] | BMOPSOFS | Wrapper (SVM) | • Maximize reliability<br>• Maximize accuracy | Sequential |
| Khan and Baig [55] | NSGA-II | Wrapper (ID3) | • True positive (TP)<br>• True Negative (TN) | Sequential |
| Xue et al. [56] | NSGA-II | Wrapper (KNN) | • Minimize classification error<br>• Minimize cardinality of the selected feature subset | Sequential |
| Xue et al. [57] | NSPSO | Filter | • Maximize relevance score<br>• Minimize redundancy score<br>using mutual information and entropy individually | Sequential |
| Xue et al. [58] | CMDPSOFS-B and NSPDOFS-A | Wrapper (KNN) | • Maximize accuracy<br>• Minimize cardinality of the selected feature subset | Sequential |
| Demir et al. [59] | MOEA/D-DYN | Wrapper (KNN) | • Minimize classification error<br>• Minimize cardinality of the selected feature subset | Sequential |
| Nguyen et al. [60] | MOEA/D | Wrapper (KNN) | • Maximize accuracy<br>• Minimize cardinality of the selected feature subset | Sequential |
| Lu et al. [61] | MOEA/D | Filter | • Maximize mutual information | Sequential |

|  |  |  |  | • Maximize pearson correlation coefficient |  |
|---|---|---|---|---|---|
| Xuan et al. [62] | M-MOEA/D | Wrapper (KNN) |  | • Maximize accuracy<br>• Minimize cardinality of the selected feature subset | Sequential |
| Luo et al. [63] | PSOMMFS | Wrapper (KNN) |  | • Maximize classification error<br>• Minimize cardinality of the selected feature subset | Sequential |
| Kozodoi et al. [64] | NSGA-II | Wrapper (rxtreme gradient boosting, logistic regression ) |  | • Maximize expected mean profit<br>• Minimize cardinality of the selected number of features | Sequential |
| Gonzalez et al. [65] | NSGA-II | Wrapper (KNN) |  | • Maximize kappa index on the training dataset<br>• Minimize kappa index on the validation dataset | Sequential |
| Deniz et al. [66] | NSPSO | Wrapper (LR) |  | • Maximize accuracy<br>• Minimize cardinality of the selected feature subset | Sequential |
| Liao et al. [27] | MOEA/D | Wrapper (KNN) |  | • Maximize classification error<br>• Minimize cardinality of the selected feature subset | Parallel |
| **Current Study** | **NSGA-II, NSPSO, MOEA/D** | **Wrapper (LR)** |  | • **Maximize AUC**<br>• **Minimize cardinality of the selected feature subset** | **Parallel** |

## 2. Literature Review

In this section, we first discuss the popular parallelization strategies using various parallel and distributed frameworks such as Message passing interface (MPI), OpenMP, Hadoop, and spark solving various MOPs. The algorithms are listed in Table 2.

Ye et al. [3] proposed a master slave NSGA-II in the message passing interface (MPI) to solve stability constrained optimal power flow (OPF). Their approach has introduced a fuzzy membership-based weight assigning technique to attain a better optimal solution. Their parallel algorithm has achieved a better acceleration effect and provided better pareto optimal solutions. Artina et al. [4] proposed an island based parallel NSGA-II in the MPI and applied it to solve small and medium-size water distribution networks. Their results indicate that their parallel models have achieved significantly better results and improved the efficacy of the exploration. Lancinskas and Zilinskas [5] designed and developed parallel NSGA-II models in both the MPI and a hybrid environment OpenMP-MPI and named the algorithms the ParNSGA/DR and ParNSGA/MPI-OMP, respectively. Authors have developed these parallel algorithms to solve the location facility problem in the multi-objective environment. Their results show that their parallel models have improved the exploration ability's efficacy than the sequential algorithms. To enhance the speedup and other hurdles such as low bandwidth, which leads to high latency, and low memory. Liu et al. [6] proposed an optimized parallel NSGA-II on the sunway taihu light system and named it swNSGA-II. The authors addressed the hurdles in the following way: (i) implemented both process level and thread level parallelism of the system based on the improved island / master-slave model. (ii) introduced a data sharing scheme based on register level communication. They have also reported swNSGA-II on the path-planning as the use-case study and reported the speedup achieved thereof. Hoang et al. [7] proposed a redistribution strategy in the island based NSGA-II to effectively perform the crossover operation on the sub population (population related to an island). Their approach has divided the population into many sub-populations, which are handled individually by a single thread. Their redistribution strategy ensures that adequate solution information is shared with all of the other sub-populations, thereby enhancing the exploration.

Oliveira et al. [8] designed a GPU based parallel NSGA-II using Compute Unified Device Architecture (CUDA) to solve the energy dispatch problem.

Table 2: Table showing the parallel implementation and the solved problems

| Authors | Algorithm | Platform (Parallel strategy) | To Solve |
|---|---|---|---|
| Ye et al. [3] | NSGA-II | MPI (Master-slave) | Stability constrained Optimal power flow |
| Artina et al. [4] | NSGA-II | MPI (Island-model) | Small and medium water distribution networks |
| Lancinskas and Zilinskas [5] | NSGA-II | MPI and OpenMPI | Location facility problem |
| Liu et al. [6] | NSGA-II | sunway taihu light system | Path-planning strategy |
| Hoang et al. [7] | NSGA-II | Multithreading (island) | Travelling salesman and Knapsack problem |
| Oliverira et al. [8] | NSGA-II | CUDA | Benchmark test functions |
| Liu et al. [9] | NSGA-II | Hadoop | Load balancing technique to optimize execution time |
| Gupta and Tan [10] | NSGA-II | CUDA | Benchmark test functions |
| Rajeswari et al. [11] | NSGA-II | Hadoop (MapReduce) | Job scheduling in the Computational grid environment. |
| Thant et al. [12] | NSGA-II | Hadoop (MapReduce based steady state NSGA-II) | Hadoop configuration optimization |
| Li et al. [13] | PSO | OpenMP, MPI | Benchmark test functions |
| Jin and Samii [15] | PSO | MPI | Wide-batch patch antenna designs |
| Nedjah et al. [16] | PSO | GPU | Benchmark test functions |
| Kondov [17] | PSO | ArFlock | High dimensional problem optimizations |
| Ding et al. [18] | PSO | Multi-threading | Benchmark test functions |
| Mussi et al. [19] | PSO | GPU | Human tracking system in the multi-view video |
| Tu and Liang [20] | PSO | Multithreading | Path planning without or with less possible obstacles |
| Mosares [21] | PSO | MPI | High dimensional engineering problems |
| Ma et al. [22] | PSO | spark | Large scale reservoir system |
| Dali and Bouamama [23] | PSO | GPU | Constraint satisfaction problems (CSPs) |
| Sato et al. [24] | MOEA/D | Multi-threading | Knapsack problem |
| Mambrini and Izzo [25] | MOEA/D | Multi-threading | Benchmark test functions |
| Souzo & pozo [26] | MOEA/D | CUDA | Knapsack problem |
| Liao et al. [27] | MOEA/D | jMetal-Multi-threading | Feature subset selection (FSS) |
| Yu et al. [28] | MOEA/D | CUDA | Benchmark test functions |
| Kantaour et al. [29] | MOEA/D | Java SE-Multi-threading | Knapsack problem |
| Sato et al. [30] | MOEA/D | Multi-threading | Knapsack problem |
| Ying et al. [31] | MOEA/D | spark | Benchmark test functions |
| **Current Study** | **NSGA-II, NSPSO, MOEA/D** | **spark (island)** | **Feature subset selection (FSS)** |

Liu et al. [9] developed a mMapReduce based NSGA-II and applied it to load-balanced based approach in the heterogeneous cloud environment. Gupta and Tan [10] proposed a scalable parallel implementation of NSGA-II in a GPU environment. Authors have applied the benchmark functions, reported the speedup, and concluded that the parallel implementations could further apply to large-scale optimization problems. Rajeswari et al. [11] proposed a MapReduce model for NSGA-II and applied it to solve job scheduling in the computational grid (CG) environment. The authors have used a fuzzy-based membership function to analyze the efficiency of the schedule. Thant et al. [12] proposed a MapReduce based steady-state NSGA-II (ssNSGA-II) and applied it to solve Hadoop configuration optimization. ssNSGA-II focused primarily on optimizing the instance resource usage and execution time of MapReduce tasks.

PSO is one of the popular algorithms due to its simplicity and power. It can be implemented in singular, multi, and many objective environments. In the literature, there are parallel PSOs approaches and their variants designed to solve various problems. Cao et al. [14] designed and proposed methodologies for the parallel PSO in various distributed and parallel frameworks. They have developed the methods for multi and many-objective environments. Jin and Samii [15] proposed a parallel PSO and finite difference time domain, PSO/FDTD, used widely in the wide-band patch antenna designs. PSO/FDTD is designed under the MPI framework. Nedjah et al. [16] designed the parallel PSO using three different methods: particle-oriented strategy, cooperation-oriented strategy, and dimension-oriented strategy. All these methods are developed in the GPU environment. The authors conducted the experiment analysis on the benchmark functions and proved that their strategy is efficient for the high dimensional problem optimizations. Kondov [17] designed a parallel PSO using the ArFlock library, which uses the MPI framework to predict the protein structure. Ding et al. [18] designed an asynchronous reinforcement learning based on the improved parallel PSO and named it APSO-BQSA. The authors have applied it to the benchmark functions. Mussi et al. [19] designed parallel PSO in GPU to solve the markerless full-body articulated human motion tracking system from the muti view video sequence acquired in a studio environment. Tu and Liang [20] designed four parallel PSO based on broadcast, star, migration, and diffusion network topologies. The authors developed this parallel version of PSO using a multithreading approach. The authors applied it to solve path planning without or with fewer possible obstacles. Mosares et al. [21] developed the parallel PSO using MPI and applied it to solve high-dimensional engineering problems. Their methods have achieved very good performance. Ma et al. [22] created a parallel PSO under the s environment and named it spark-based parallel PSO (SPPSO) via cloud computing. SPPSO is designed to solve large-scale reservoir systems. The simulation experiments are conducted on the eight reservoir systems located in China. SPPSO has achieved better stability, performance, efficiency than sequential dynamic programming. Dali and Bouamama [23] developed the parallel version of PSO in GPU to solve constraint satisfaction problems (CSPs). They developed two novel approaches: (i) parallel GPU for max CSPs and (ii) GPU distributed PSO for max-CSPs. Recently, Li et al. [13] proposed generation-level parallelism in the PSO and named it P3SO, inspired by the pipeline-based approach. The generation level parallelism works as follows: few populations still perform the current generation while others jump into the next generation. This is the first-of-its-kind study, where parallelism is performed at the generation level. So far the parallelism is performed at the population level, where the population is divided into few sub populations and runs in parallel. But, in the generation level parallelism, a few of the solutions evolve at the current generation while few others evolve in the next generation.

Parallel MOEA/D on parallel and distributed frameworks may attain good speedup but may result in poor uniformity and diversity in achieving the pareto optimal solution set. Hence, Sato et al. [24] proposed a many-core based MOEA/D with the standard migration rule, where the overlapping zone is computed, and then the pareto optimal solution set is selected. Authors have compared with a non-migration based MOEA/D, and the results indicated that the MOEA/D with migration rule attained better solutions to solve the Knapsack problem. Mambrini and Izzo [25] designed and developed a parallel decomposition (PaDe), an asynchronous algorithm that uses the multithreading framework and solves various benchmark solutions. Souza and Pozo [26] designed a hybrid Evolutionary algorithm, viz., MOEA/D and ant colony optimization (ACO), parallel MOEA/D-ACO, using the CUDA in GPU environment. Authors have applied the parallel MOEA/D-ACO to solve the multi-objective knapsack and multi-objective traveling salesman problems. Liao et al. [27] proposed a parallel version of the MOEA/D-STAT, a modified version of MOEA/D with multiple reference points, and MOEA/D and applied it to solve FSS. Authors have designed the algorithms using the jMetal platform, which leverages them to implement in multi-threaded based parallelization. Authors have developed both the master-slave and island based models. Both these methods used KNN as the classifier. Yu et al. [28] proposed a CUDA-based version of MOEA/D and applied it to solve various benchmark functions. The authors reported that their parallel strategy had attained a good speedup. Kantaour et al. [29] proposed an MOEA called parallel criterion based partitioning MOEA (PCPMOEA) via multithreading in the Java SE platform. Authors have applied PCPMOEA to solve the knapsack problem. Sato et al. [30] proposed multithreading based on many cores applicable MOEA/D method, which is suitable for effectively dealing with the sparse area's problems. They have devised a new method that is used to switch between the reference points during the update interval. They further applied the algorithm to solve the knapsack problem. Ying et al. [31] designed two parallel and distributed MOEA/D under the spark environment where (i) parallel model-I evolves the population entirely on spark and (ii) parallel model-II evolves the population partially on the spark. They applied it to a few benchmark functions to prove the efficacy of the proposed parallel approaches.

Eventhough there are several variants of NSGA-II, NSPSO and MOEA/D developed for FSS (refer to Table 1) yet all are sequential in nature. The information presented in Table 1 is as follows: information related to the objectives chosen, whether these objectives are maximized/minimized and which classifier is considered. Our current approach is also included to show the difference in the approach, type and nature of the algorithm. Our current approach is different from the [46,53,58,60,62,66] studies, in that all of them considered accuracy and cardinality as the objective functions whereas we considered AUC and cardinality. AUC is preferred over accuracy because it is more meaningful than latter in unbalanced datasets. Also, none of the studies mentioned in Table 1 except Liao et al [27] comes close to our study in that they also proposed a parallel scheme for FSS. However, our study is distinct from Liao et al [27] in the following way:

- The major difference is with respect to parallel model. Liao et al. [27] used jMetal multithreading framework and did not test their approach on big datasets. However, the current work is developed under the spark environment and it is tested on big and high dimensional datasets.
- They proposed parallel MOEA/D, whereas, the current work proposed parallel versions of NSGA-II, NSPSO and MOEA/D.
- Liao et al. [27] adopted elitist based migration policy, where the best solution of one data island is shared with the other data islands. Thus, only the top solutions are considered may result in entrapment in local minima. However, our work used non-dominated sorting (NS) principle based migration based policy / rule. Here the evolved populations are collected from different data islands and then NS principle is applied on them. This strategy filters out the poor solutions and only the top non-dominated solutions get to survive.

The major focus of the current research work is as follows:
- Current parallel island model is designed under spark environment to solve FSS problem. Moreover, to the best of the knowledge, little or no work is reported solving FSS for big datasets in the bi-objective environment using MOEA based wrappers under spark environment.
- Further, the parallel model is designed with a migration strategy which helps in selecting top few solutions evolved from various islands.

# 3. Background of the metaheuristics employed

This section briefly discusses the background of the vital concepts and the overall metaheuristics employed in the current research work.

## 3.1 Non-dominated solutions

Multi-objective optimization problems (MOPs) involve objective functions. It is difficult to choose the best solution when there are several objective functions as there is a chance of trade-off/conflict among them. Assume that we have $\lambda$ objective functions. A solution $x_1$ is said to be dominating the $x_2$ if and only if one of the following conditions is met:
   a) Solution $x_1$ is strictly better than solution $x_2$ in at least one among the $\lambda$ objective functions (or) $f_j(x_1)$ is better than the $f_j(x_2)$ $\exists\ j \in \{1,2,..\lambda\}$; where $f_j$ is the $j^{th}$ objective function and $\lambda$ is the number of objective functions.
   b) Solution $x_1$ is not worse than the solution $x_2$ in all of the $\lambda$ objective functions (or) $f_j(x_1)$ is better than the $f_j(x_2)$ $\forall\ j \in \{1,2,..\lambda\}$; where $f_j$ is the $j^{th}$ objective function and $\lambda$ is the number of objective functions.

## 3.2 Pareto Optimal Front

All the solutions which are non-dominated to each other are assigned with the same domination rank. The lower the domination rank, the better the solutions and the same domination rank can be assigned to multiple solutions. The solutions having the same domination rank forms a solution set which are non-dominated to one another thus forming a pareto optimal front.

## 3.3 Crowding distance

Crowding distance determines the density of the solutions. It is calculated as follows: sort the solutions in the ascending order for each objective function. The boundary solutions highest and lowest solution are assigned with inifinite crowding distance values to make their selection mandatory. The rest of the solutions are assigned with the mean distance of its two neighbouring solutions i.e., normalized objective score. This is repeated for each objective function. Generally, the final crowding distance value is the aggregation of the combined individual crowding distance scores. It is used to rank the solutions within the pareto optimal front.

## 3.4 NSGA-II

NSGA-II, proposed by Deb et al. [32], has proven one of the most efficient algorithms in the multi-objective environment. This metaheuristic works on the non-dominated sorting principle and a crowding distance to dispose of the solutions into various pareto optimal fronts.
The algorithm starts with initializing the population (*P*) of size *N*, followed by the objective function evaluations on thus initialized population. Now, the population is subjected to the standard one-point crossover and polynomial mutation resulting in the generation of the offspring population. Then, the offspring population is subjected to fitness score evaluation. The offspring population is now combined with the population results in a total size of *2N* and then applied with a non-dominated sorting phase, where only the top *N* solutions survive and act as the population for the next iteration. This process, subjected to the crossover and polynomial mutation, is repeated for *maxIter* number of iterations.

## 3.5 NSPSO

NSPSO is the modified version of the PSO and is made suitable for the multi-objective optimization proposed by Li [33]. NSPSO is designed by incorporating the PSO search mechanism with the multi-objective environment principles.

The algorithm starts with initializing the population $P$ of size $N$ with $d$ dimensions; followed by the objective function evaluations on the initialized population. Let the personal best of the solution $i$, is $Pbest_i = (pbest_{i1}, pbest_{i2}, ..., pbest_{id})$. This indicates the best position that yielded the highest fitness value for the i$^{th}$ solution. Also, the position of the best solution is stored in the $Gbest_g = (Gbest_{g1}, pbest_{g2}, ..., pbest_{gd})$. Then the population undergoes the non-dominated sorting principle, thereby assigning the solutions to a pareto optimal front. Then, based on the crowding distance, the solutions are ranked within the front. Now, each particle of the solution undergoes the *velocity update* (refer to Eq. (1)) followed by the *position update* (refer to Eq. (2)) resulting in a real encoded vector. This undergoes the sigmoid (refer to Eq. (3)) for generation of the offspring population. Then, the non-dominated sorting phase, where only the top $N$ solutions survive out of the total 2N solutions, is performed. The surived solutions acts as the population for the next iteration. This process, which is subjected to the position update and velocity update, is repeated for *maxIter* number of iterations.

$$v_{ij} = w * v_{ij} + c_1 * r_1 * (Pbest_{ij} - P_{ij}) + c_2 * r_2 * (Gbest_{gj} - P_{gj}) \qquad (1)$$

where, $ij$ represents the $j^{th}$ member of the $i^{th}$ solution, w is the inertia weight, $c_1, c_2$ are the control parameters (positive), r1, r2 are the two random numbers that lie in the range [0,1].

$$P_{ij} = P_{ij} + \Omega * v_{ij} \qquad (2)$$

where, $\Omega$ is the construction factor that constricts and controls the velocity magnitude.

$$P_{ij} = \begin{cases} 1, & if \ rand(0,1) < sigmoid(P_{ij}) \\ 0, & else \end{cases} \qquad (3)$$

## 3.6 MOEA/D

In this section, we will first discuss briefly the decomposition method used in the MOEA/D. Then, we briefly discuss the MOEA/D algorithm.

MOEA/D, proposed by Zhang and Li [34], is a decomposition based algorithm where the MOPs are sub-divided into multiple single objective functions and each of the objective functions is optimized simultaneously. Such decomposition strategies are of three types: (i) weighted sum approach (ii) tchebycheff approach and (iii) boundary crossing method. In the current study, techebycheff approach is adopted because it is too popular and well-proven for multi objective problems. The techebycheff approach invokes scalar optimization as presented in Eq. (4). Note that, in this approach, a different optimal solution can be obtained by altering the weight vector. The advantage of this approach is that the optimal solutions are evenly distributed along the pareto front. Also, there is no need to compute the derivative of the aggregation function unlike the other two approaches.

$$minimize \ g^{te}(x|\lambda, z^*) = \max_{1 \leq i \leq m} \{\lambda_i | f_i(x) - z_i^*|\} \qquad (4)$$

where, $z^* = (z_1^*, ...., z_\lambda^*)$ acts as a reference point for each of the objective $i = 1,2,..,m$. For each optimal solution $x^*$, there exists a weight vector, $\lambda$.

MOEA/D, a decomposition-based method, works on the principle of decomposing the MOP into a number of SOPs and optimizing independently through decomposition methods. MOEA/D maintains an archive of the population, EF, to store the non-dominated solutions found in the search. The uniform set of vectors is also initialized, which helps in obtaining the neighborhood solutions. Based on the Euclidean distance, assign the T closest vectors to each weight vector. These neighborhood solutions are updated in every generation by following Eq. (5).

$$if\ g^{te}(y|\lambda_j, z^*) \leq g^{te}(x_j|\lambda_j, z^*)\ then\ replace\ x_j = y \tag{5}$$

where, $x_j \in \text{NB}(x)$, neighbourhood solution of $x$.

MOEA/D starts with initializing the population randomly. Then, the population is subjected to the standard one point crossover and the polynomial mutation, resulting in the generation of the offspring population. The objective function scores are also computed upon the generated offspring population. Based on the decomposition strategy, update the neighborhood solution set. Later the archive population is also updated with the non-dominated solutions by combining the current population with the evolved population in the current iteration. This process is continued for *maxIter,* the number of iterations.

## 4. Proposed Island Model under Apache spark

This section starts with a discussion about the objective functions overview of the parallel mechanism that is employed for all the algorithms. The discussion continues with the P-NSGA-II-IS, P-NSPSO-IS, P-MOEA/D-IS, and the complexity of the algorithms.

### 4.1 Objective Functions

Strictly speaking, the FSS falls under the standard combinatorial optimization problem. The critical task here is to achieve the highest possible AUC by minimizemaximi the number of selected features. Hence, FSS is considered as the bi-objective optimization problem where:

(i) Cardinality : It is given by the number of selected features in the given solution. Mathematically, it is the ratio of the cardinality of the selected feature subset to the total number of features.

$$f_1 = Cardinality = \frac{\#SelectedFeatureSubset}{\text{Total No. of Features}} \tag{6}$$

(ii) Area under the ROC Curve (AUC) : underlying classifier calculates the AUC (refer to Eq. (7) achieved by those selected feature subset. AUC is chosen because it is proven to be robust measure than accuracy. It is defined as the mean of specificity and sensitivity.

$$f_2 = AUC = \frac{Sensitivity + Specificity}{2} \tag{7}$$

The mathematical formulation of the bi-objective optimization problem is defined in Eq. (8). Here, we minimize the cardinality, which is calculated from Eq. (6) and maximize the AUC, which is computed using Eq. (7).

$$Minimize\ f_1\ and \tag{8}$$
$$Maximize\ f_2$$

In this work, LR is employed as the classifier for the proposed FSS. LR is chosen because it is fast to train and is nonparametric. It does not make any assumptions about the errors or variables, takes less time to converge, and has no hyperparameters to fine-tune.

## 4.2 Encoding scheme of the population

Fig. 1 depicts the population schema that was maintained for all approaches in this work. The population consists of a solution of size *N*. Each solution has two different fields: (i) Key field having the unique id information to identify the solution uniquely, (ii) The Value field has the following subfields: (a) binary vector: which is of length number of features, *nfeat* is a binary vector where the presence of '0' says that a particular feature is not selected, and '1' says that the particular feature is selected. (b) selected features store the selected features as a vector. (c) trained model coefficients: as we know that in the wrapper methods, a model is chosen to evaluate the solution's performance. Thus trained model coefficients are stored in this sub-field. The main reason for storing them is to use them in the test phase. (d) AUC: The trained model results an AUC for each solution. That information is stored in the sub-field. (e) Cardinality : The selected features' length is calculated and stored in this sub-field. This kind of schema makes sure to preserve the solution-related information in a single space.

| Key | Value : Solution Vector |
|---|---|
| $Key_{\{1\}}$ | $<$ Binary Vector$_1$, selected Features$_1$, trainedModelCoeffcients$_1$, AUC Score$_1$, FSC Score$_1$ $>$ |
| $Key_{\{2\}}$ | $<$ Binary Vector$_2$, selected Features$_2$, trainedModelCoeffcients$_2$, AUC Score$_2$, FSC Score$_2$ $>$ |
| ... | ... |
| $Key_{\{N\}}$ | $<$ Binary Vector$_N$, selected Features$_N$, trainedModelCoeffcients$_N$, AUC Score$_N$, FSC Score$_N$ $>$ |

Fig. 1: Schema of the population

Table 3: List of Notations used in the current research study

| Notation | Meant for |
|---|---|
| $N$ | population size |
| *localN* | sub-population size |
| *mMig* | maximum number of migrations |
| *mGen* | maximum number of generations |
| $P$ | population |
| *Xtrain* | training dataset |
| *Xtest* | test dataset |
| *pc* | crossover probability |
| *pm* | mutation probability |
| *w* | inertia weight |
| $c_1, c_2$ | predefined constants of PSO |
| $T$ | number of neighbourhood solutions |

## 4.3 Proposed Generic framework for the parallel island model

This subsection will discuss the generic framework for the bi-objective algorithms designed under the spark with the NS-based migration policy. The generic framework flowchart is depicted in Fig.2, and the schematic is depicted in Fig.3. The list of notations frequently used in the current study is given in Table 3. In the current island model, the data is initially divided into k-number of partitions, referred to as the data islands. The island model is divided into three main stages, namely, (i) Population Initialization, (ii) Training Phase (iii) Test Phase.

**Stage-I: Population Initialization**

The first stage, where the population is randomly initialized of size $N$. It follows the encoding scheme as discussed in section 4.2. The initialized population acts as the global population. Once the population is initialized, it is divided into k-number of randomly selected overlapped sub-populations each of size *localN*. In addition, all the required control parameters are broadcasted to all the data islands. This ensures that each of the islands gets a diverse initial sub-population. Thus created sub-population serves as the randomly initialized population of each island and is referred to as the local population in this work.

**Stage-II: Training Phase**

Training phases mainly involve:
    (a) The evaluation of the initial sub-population. This is done parallel on the worker nodes.
    (b) Evolving of the population, which is done for *mGen* maximum number of generations. This is also done in parallel on the worker nodes.
    (c) Upon completing the *mGen* number of generations, each data island results in an evolved population. Here, the migration rule/policy filters out poor non-dominated solutions. This migration rule is applied only at the driver node. The major objective for the invocation of migraton rule is that the information sharing happens across the evolved population from various data islands.

As a first step, for each solution in the local population, the selected feature subset is obtained from the binary encoded vector of the initialized population. Then AUC evaluation, one of the objective functions, is done for each of the solutions of the local population. Note that the trained LR model gives the model coefficients. These LR model coefficients are stored for each of the solutions. After that, the cardinality of a solution, another objective function, is obtained by counting the number of selected features. It is important to note that the same scheme of evaluation, the AUC and the cardinality are being used throughout the process whenever the objective score calculation is done. Once the AUC and cardinality are computed on the initialized population, the population evolving process begins.

    Based on the metaheuristic, the underlying heuristics are applied to the local population resulting in the offspring population. Now, LR evaluates the AUC based on the selected feature subset obtained from the binary encoded vector of the offspring population. The cardinality is also obtained by a number of selected features in each solution of the offspring population. Once this is completed, the Non-dominated sorting (NS) principle is applied in the case of pareto-optimal based bi-objective metaheuristic (i.e., here in the case of NSGA-II, NSPSO) and in the case of decomposition based bi-objective metaheuristic (i.e., here in the case of MOEA/D). In either of the cases, it acts as the *selection* operator and results in the formation of a new local population, which serves as the parent population for the next iteration. This employing of heuristics and the selection phase is continued for the *mGen* number of generations. This is done individually for all the data islands, and the local population is evolved parallel across the data islands.

    Once after completing the mGen number of generations, each data island results in an evolved population. Suppose the size of the evolved population is of *localN* from each of the k-data islands, which results in *k\*localN* number of solutions. All these evolved populations are collected back to the driver where the migration policy is applied. The major objective for the invocation of migration rule is that the information sharing happens across the evolved population from all data islands. Here, the solutions evolved from a particular island may not be the best in the context of other island. Hence, blindly picking the best solutions may lead to entrapment in the local minima. Hence, to maintain the disparity in the evolved population the we adopted NS principle based migration rule. Employing NS

based migration policy result in the pareto-fronts. Then the *top-N* solutions are selected, and this population is again sub-divided into k sub-populations of *size N* each and is sent back to the data islands. Instead of picking the best solutions blindly from each island now the non-dominated set of solutions is obtained. Then the training phase is again executed for *mGen* times. This training phase and then employing the migration policy completes one migration. Thus selected population serves as the global population for the next migration. This whole training phase and employing migration policy is executed for a *defined* number of migrations, say *mM*ig.

**Stage-III Test phase**

Upon the completion of *mMig* migrations, results in the evolved population, which is evaluated on the test dataset in the test phase. The LR coefficients of each of the solutions are obtained from the $trainedModelCoeffcients$ column of each of the solution (refer to Fig. 1). This completes the test phase and then the test AUC and cardinality are reported.

Note that all the algorithms have the same population initialization, way of calculating the two objective functions viz., (i) AUC score evaluation, (ii) cardinality of the feature subset calculation, NS-based migration rule, and test phase. Hence, the discussion on these is obviated. Also, we only focused majorly on the algorithmic specific changes in the subsequent sections.

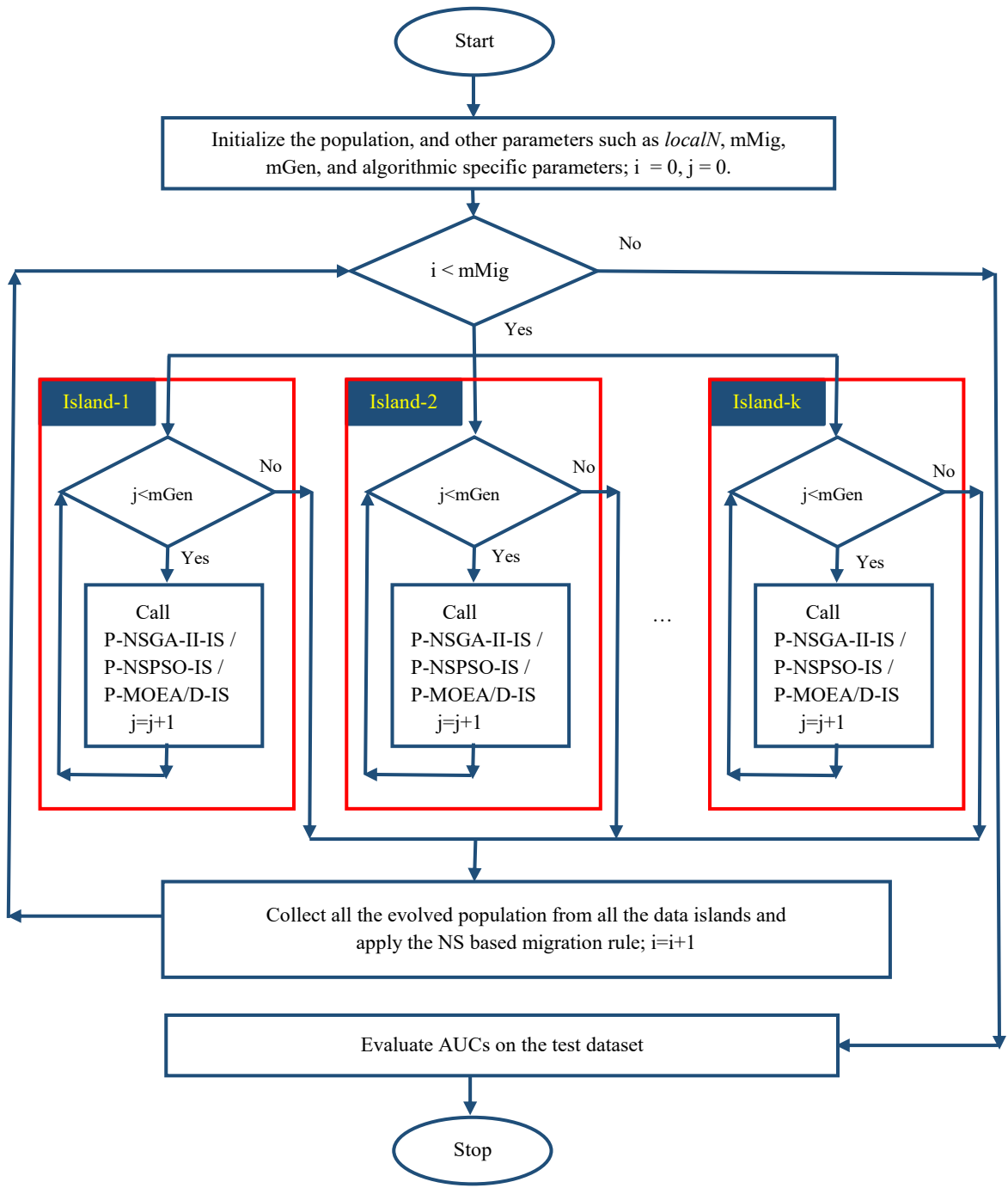

Fig. 2: Generic flowchart for the bi-objective algorithms in the island model based wrapper

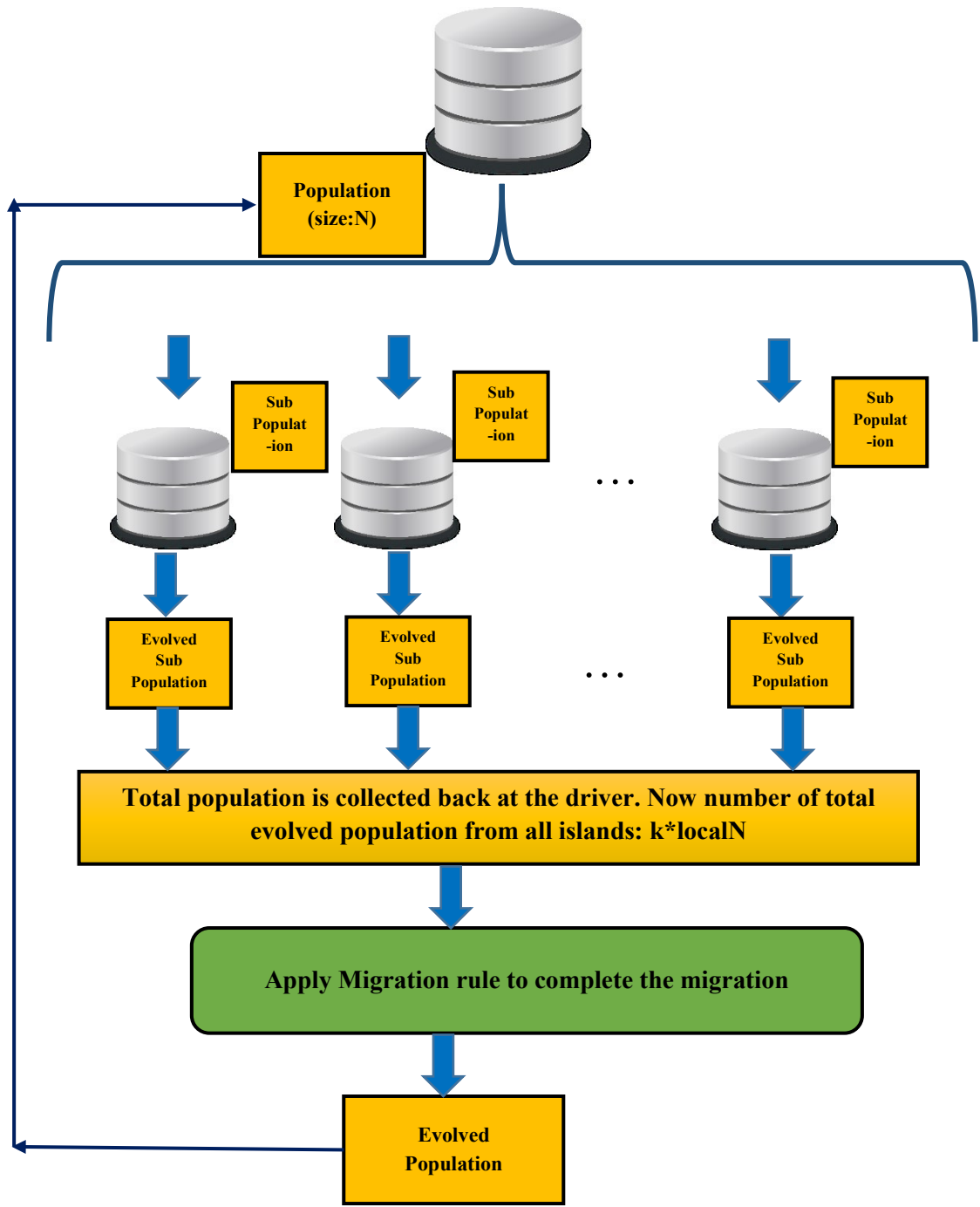

Fig. 3: Generic schematic diagram for the bi-objective algorithms in the island model based wrapper

## 4.4 Parallel NSGA-II island model

This subsection deals with the P-NSGA-II-IS model, which follows the generic parallel framework as discussed in section 4.3. P-NSGA-II-IS starts with initializing the population (see Algorithm 1) and follows the structure as depicted in Fig.1. All the required control parameters, namely the *mGen, mMig, pc, pm, N, localN* (refer to Table 3) are initialized accordingly. Then the *P-NSGA-II-IS-Mapper* is called, begins the training phase, and then shift the control to the workers (see Algorithm 2).

*Algorithm 1: P-NSGA-II-IS Driver Algorithm*

> **Input:** N, localN, P, Xtrain, Xtest, mMig, mGen, pc, pm
> **Output:** P : population evolved after mMig migrations
> 1: i ← 0;
> 2: P ← Randomly Initialize the population
> 3: Divide the Xtrain into k number of islands
> 4: whille ( i < mMig){
> 5:     Call P-NSGA-II-IS-Mapper (namely, Algorithm 2)
> 6:     **Migration Rule:**
>         a) Total population ← Collect all the evolved populations from all the islands
>         b) Apply Non dominating sorting principle on the Total population
>         c) Arrange the solutions as per the Crowding distance with in the front.
>         c) P ← Consider top solutions of size N.
> 7:     i=i+1 }
> 8: coef ← collect coefficients of all the solutions
> 9: Evaluate Test AUCs using coef of the evolved population P
> 10: return P, Test AUCs

**Training phase**

Once the data island gets the initial sub-population (here, local population pertained to data island), the objective function scores are calculated for each solution. The cardinality is calculated, and the LR evaluates the AUC on the selected feature subset obtained by each solution of the sub-population (step 4 train-update phase in Algorithm 2). Then the NSGA-II heuristics viz., one-point crossover, and polynomial mutation are applied on the sub-population results in the offspring population (steps 7,8 in Algorithm 2). Then the objective function scores are calculated accordingly (step 10 in Algorithm 2). Thereafter, the offspring population is combined with the parent population, resulting in size *2N*. Then the Non-dominated sorting (*NS*) principle (steps 11-14 in Algorithm 2) is applied, and then *top-N* solutions out of total *2N* are selected. This population serves as the parent population for the next generation. This process is repeated for the *mGen* number of generations. The above process is done by the *workers* simultaneously, thereby leveraging all the sub-populations (pertaining to *k*-data islands) to evolve independently.

The above step in the training phase results in an evolved population of size N from each k-data islands, resulting in a *k*N* number of solutions collected back to the *driver*. After that, the migration policy (step 6 in Algorithm 1) is applied and involves the invocation of the Non-dominated sorting (*NS*) principle, and thereby resulting in the different pareto-fronts.

This training phase followed by employing the migration policy completes one migration. This whole training phase and employing migration policy is executed for *mMig* number of migrations.

Once the *mMig* is completed, then the test phase begins (step 9 in Algorithm 1), where the test AUCs are computed (refer to Section 4.3).

*Algorithm 2: P-NSGA-II-IS Worker Algorithm*

```
Input: localN, mGen, pc, pm
Output: P: population evolved after mGen, max generations
1: function P-NSGA-II-IS-Mapper{
2:     k ← 0;
3:     localP ← Randomly pick the localN size of subset from P    // local population
4:     train-and-update phase:
         a) Evaluate LR on localP.BinaryEncodedVector .
         b) Store thus computed coefficients and compute AUC, cardinality .
         c) update the objective function scores and coefficients in the appropriate fields.
5:     whille ( k < mGen){
6:          localPVector ← localP.BinaryEncodedVector
7:          crossoveredVector ← perform One point Crossover on localPVector
8:          mutatedVector ← perform Polynomial Mutation on mutatedVector
9:          Create localOffspringP using mutatedVectors      // local offspring population
10:         Perform train-and-update phase on localOffspringP
11:         combinedLocalP ← localOffspringP U localP
12:         fronts ← perform the Non-dominated sorting principle on the combinedLocalP
13:         fronts ← compute Crowding Distance for each front and update the order
14:         localP ← consider the N number of solutions as per the order in the fronts.
15:         k=k+1 }
16: return localP}
```

## 4.5 Parallel NSPSO island model

P-NSPSO-IS starts with initializing the population (refer to Algorithm 3) and follows the structure as depicted in Fig.1. All the required control parameters, namely the *mGen, mMig, w, c1, c2, N, localN* (refer to Table 3) are initialized accordingly. Then the training phase is started, and now the control is shifted to the workers (refer to Algorithm 4).

**Training phase**

Once the data island gets the initial sub-population, the velocity is also initialized accordingly (steps 3-4 in Algorithm 4). Now, the objective function scores are calculated, i.e., LR evaluates the selected feature subset obtained from the binary vector and evaluates the AUC for the entire sub-population (step 5 in Algorithm 4). Then, the local best and the global best solutions are obtained (steps 8,9 in Algorithm 4). Now, the NSPSO heuristics viz., velocity update, and position are done (steps 10-12 in Algorithm 4) and are applied to the offspring population's sub-population results. Then the objective function scores are calculated accordingly (step 13 in Algorithm 4). After that, the offspring population is combined with the parent population, resulting in size *2N*. Then the Non-dominated sorting (*NS*) principle (steps 14-18 in Algorithm 4) is applied, and then *top-N* solutions out of a total *of 2N* solutions are selected. This population serves as the parent population for the next generation. This process is repeated for *the mGen* number of generations. The above process is also done by the workers simultaneously, thereby leveraging all the sub-populations to evolve independently.

*Algorithm 3: P-NSPSO-II-IS Driver Algorithm*

**Input:** N, localN, P, Xtrain, Xtest, mMig, mGen, w,c1,c2
**Output:** P : population evolved after mMig migrations
1: i ← 0;
2: P ← Randomly Initialize the population
3: Divide the Xtrain into k number of islands
4: whille ( i < mMig){
5:     Call P-NSPSO-IS-Mapper (namely, Algorithm 4)
6:     **Migration Rule:**
        a) Total population ← Collect all the evolved populations from all the islands
        b) Apply Non dominating sorting principle on the Total population
        c) Arrange the solutions as per the Crowding distance with in the front.
        c) P ← Consider top solutions of size N.
7:     i=i+1 }
8: coef ← collect coefficients of all the solutions
9: Evaluate Test AUCs using coef of the evolved population P
10: return P, Test AUCs

*Algorithm 4: P-NSPSO-II-IS Worker Algorithm*

**Input:** localN, mGen, w, c1, c2
**Output:** P: population evolved after mGen, max generations
1: function P-NSPSO-IS-Mapper{
2:     k ← 0;
3:     localP ← Randomly pick the localN size of subset from P    // local population
4:     velocity ← Initialize the velocity vector for the local population
5:     **train-and-update phase:**
        a) Evaluate LR on localP.BinaryEncodedVector .
        b) Store thus computed coefficients and compute AUC, cardinality .
        c) update the objective function scores and coefficients in the appropriate fields.
6:     while ( k < mGen){
7:         localPVector ← localP.BinaryEncodedVector
8:         gbest ← get the global best solution
9:         lbestPop ← get the local best solutions
10:        velocity ← update the velocity for each solution using gbest (Step 8) and
           lbestPop (step 9)( as per Eq. (1))
11:        positionVectors ← update the position for each solution using velocity
           obtained from (Step 10) ( as per Eq. (2) & (3))
12:        Create localOffspringP using positionVectors        // local offspring population
13:        Perform train-and-update phase on localOffspringP
14:        combinedLocalP ← localOffspringP U localP
15:        fronts ← perform the Non-dominated sorting principle on the combinedLocalP
16:        fronts ← compute Crowding Distance for each front and update the order
17:        localP ← consider the N number of solutions as per the order in the fronts.
18:        k=k+1 }
19: return localP}

After completing the training phase, from each data island an evolved population of size *localN* is obtained. As there are k data islands which result in total *k*localN* number of solutions are collected back to the driver. After that, the migration policy (step 6 in Algorithm 3) is applied and involves the invocation of the Non-dominated sorting (*NS*) principle, and thereby resulting in the different pareto-fronts. This completes one migration. This whole process is executed for *mMig* number of migrations.

Once the *mMig* migrations *are* completed, then the test phase begins (step 9 in Algorithm 1), where the test AUCs are computed after that reported (refer to Section 4.3).

## 4.6 Parallel modified MOEA/D island model

In this section, we discussed the modified MOEA/D where the NS principle is incorporated via migration policy while paralleizing it. P-MOEA/D-IS starts with initializing the population (refer to Algorithm 5) and follows the structure depicted in Fig.1. All the required control parameters, namely the *mGen, mMig, pc, pm, N, localN* (refer to Table 3) are initialized accordingly. Then the training phase is started, and then the control is shifted to the workers (refer to Algorithm 6).

*Algorithm 5: P-MOEA/D-IS Driver Algorithm*

**Input:** N, localN, P, Xtrain, Xtest, mMig, mGen, pc, pm, T
**Output:** P : population evolved after mMig migrations
1: i ← 0;
2: P ← Randomly Initialize the population
3: Divide the Xtrain into k number of islands
4:  whille ( i < mMig){
5:     Call P-MOEA/D-IS-Mapper (namely, Algorithm 6)
6:     **Migration Rule:**
       a) Total population ← Collect all the evolved populations from all the islands
       b) Apply Non dominating sorting principle on the Total population
       c) Arrange the solutions as per the Crowding distance with in the front.
       c) P← Consider top solutions of size ps.
7:     i=i+1 }
8: coef ← collect coefficients of all the solutions
9: Evaluate Test AUCs using coef of the evolved population P
10: return P, Test AUCs

**Training phase**

Once the data island gets the initial sub-population, the external archive population and the optimal solution are initialized accordingly. Also, the neighbourhood solutions are also calculated as per the euclidian distance (step 2-6 Algorithm 6). Now, the objective function scores, i.e., cardinality , are computed, and LR evaluates the AUC for the population (step 7 Algorithm 6). Then, the MOEA/D heuristics viz., one-point crossover, and polynomial mutation are applied on the sub-population results the offspring population (step 9-12 Algorithm 6). Then the objective function scores are calculated accordingly. Thereafter, the offspring population is combined with the parent population, resulting in size *2N*. Then the Non-dominated sorting (*NS*) principle is applied, filters the *top-N* solutions, serves as the parent population for the next generation. This process is repeated for *the mGen* number of generations. Here, the above process is done in the workers simultaneously, thereby leveraging all the sub-populations to evolve independently.

After completing the training phase, obtains *k\*localN* number of solutions collected back to the driver. After that, the migration policy is applied (step 6 Algorithm 5) involves the invocation of the Non-dominated sorting (*NS*) principle, thereby resulting in the different pareto-fronts. This training phase and then employing the migration policy completes one migration. This whole training phase and employing migration policy is executed for *mMig* number of migrations.

Once the *mMig* migrations are completed, the test phase begins (step 9 Algorithm 1), where the test AUCs are computed after that reported (refer to Section 4.3).

*Algorithm 6: P-MOEA/D-IS Worker Algorithm*

*Input: localN, mGen, pc, pm, T*
*Output: EP: updated archive population obtained after mGen, max generations*
1: function P-MOEA/D-IS-Mapper{
2:     k ← 0;
3:     localP ← Randomly pick the localN size of subset from P    // local population
4:     EP ← ϕ                              // Initialize archive population
5:     NB ← Calculate T closest neighbourhood solutions using euclidian distance
6:     $z^*$ ← initialize the optimal fitness value for each objective function that can be attained.
7:     **train-and-update phase:**
        a) Evaluate LR on localP.BinaryEncodedVector .
        b) Store thus computed coefficients and compute AUC, cardinality .
        c) update the objective function scores and coefficients in the appropriate fields.
8:     whille ( k < mGen){
9:         localPVector ← localP.BinaryEncodedVector
10:        crossoveredVector ← perform One point Crossover on localPVector
11:        mutatedVector ← perform Polynomial Mutation on mutatedVecto
12:        Create localOffspringP using mutatedVectors     // local offspring population
13:        Update the $z^*$
14:        NB ← Compute Function values (refer to Eq. (4)) and
               update the neighbourhood solutions (refer to Eq. (5))
15:        Perform train-and-update phase on localOffspringP
16:        EP ← EP U localP
17:        fronts ← perform the Non-dominated sorting principle on the EP
18:        fronts ← compute Crowding Distance for each front and update the order
19:        EP ← consider only the non-dominated solutions
20:        k=k+1 }
21: return EP}

## 5 Dataset Description

The benchmark datasets and their meta-information is presented in Table 4. All the datasets are binary classification datasets. The Epsilon dataset is from the LIBSVM [40] binary dataset repository, the IEEE malware dataset is from the IEEE data port [39], genomic datasets viz., OVA_Omentum and OVA_Uterus datasets are from the OpenML repository [40], and the DDoS dataset is from the DDoS

UNB repository [41]. All the datasets have numerical features except the DDoS dataset. The DDoS dataset is of a highly imbalanced type. Hence, we have adopted the data balancing technique viz., random oversampling (ROS).

Table 4: Description of the benchmark datasets

| Name of the Dataset | # Objects | # Features | # Classes | Size of the Dataset |
|---|---|---|---|---|
| Epsilon [40] | 5,00,000 | 2000 | 2 | 10.8 GB |
| IEEE Malware [39] | 15,00,000 | 1000 | 2 | 3.2 GB |
| OVA_Omentum [40] | 1584 | 10,935 | 2 | 108.3 MB |
| OVA_Uterus [40] | 1584 | 10,935 | 2 | 108.3 MB |
| DDoS [41] (after preprocessing) | 77,95,413 | 88 | 2 | 18.2 GB |

# 6 Results and discussion

The train-test set split ratio is maintained as 80:20 with stratified random sampling for all the datasets. After rigorous experimentation, the hyperparameters for all the algorithms and all the datasets are fixed and the values presented in Table 5. The population size, sub-population size per island, and the maximum number of generations per migration are also considered as hyperparameters. All the algorithms are maintained with the same population size, subpopulation size per island, which is taken as half of the population size and the number of generations per migration. The number of migrations is fixed to be 1(one). All the experiments are repeated for 20 runs to nullify the impact of the random seed on the results.

## 6.1 Environmental Setup

All the experiments are conducted in a spark-HDFS cluster with spark version 2.4.4 [38] and Hadoop version 2.7, having one master node and four worker nodes with 32 GB RAM with Intel i5 8$^{th}$ generation.

Table 5: Hyperparameters for all the approaches

| Dataset | #population (#subpopul-ation per data island) | Total Generations (#maxGen per migration) | P-NSGA-II-IS | | P-NSPSO-IS | | | P-MOEA/D-IS | |
|---|---|---|---|---|---|---|---|---|---|
| | | | pc | pm | w | c1 | c2 | pc | pm |
| Epsilon | 50 (25) | 50 (25) | 0.96 | 0.03 | 0.9 | 0.7 | 1.3 | 0.95 | 0.10 |
| IEEE Malware | 50 (25) | 50 (25) | 0.91 | 0.05 | 0.1 | 0.6 | 1.4 | 0.95 | 0.10 |
| OVA_Omentum | 300 (200) | 100 (50) | 0.99 | 0.0004 | 0.9 | 0.3 | 1.7 | 0.99 | 0.006 |
| OVA_Uterus | 300 (200) | 100 (50) | 0.99 | 0.0007 | 0.9 | 0.5 | 1.5 | 0.99 | 0.004 |
| DDoS | 20 (10) | 10 (5) | 0.95 | 0.05 | 0.9 | 0.4 | 1.6 | 0.95 | 0.05 |

## 6.2 Experiment analysis based on the Empirical attainment function

This section discusses the performance analysis of the proposed parallel algorithms on various datasets using the Empirical attainment function (EAF) plots [68, 69] and implemented using an R package [70]. This is very important because each run produces a non-dominated set approximating the pareto optimal front. This analysis emphasizes the summary of the outcomes of multiple runs of the MOEAs and gives insights into where and whether an algorithm is better than another. It can be seen as a distribution of the quality solutions after running a particular algorithm for a certain amount of runs. Suppose for a particular dataset, where the population size is $N$ and executed for $k$ runs, then the number of the solution generated in the k runs would be $N * k$. All these solutions are used to plot EAFs. In the current study, each algorithm is executed for 20 runs, and the population size ($N$) is presented for a particular dataset in Table 4. This process is followed as same for all the algorithms.

All the EAF plots for all the algorithms and datasets are depicted in Appendix. The EAF function plots of Epsilon is depicted in the Fig.A.1, IEEE malware dataset in the Fig.A.2, OVA_Omentum in the Fig.A.3, Uterus in the Fig.A.4, and DDoS dataset in the Fig.A.5. For each dataset, the EAFs corresponding to P-NSGA-II-IS, P-NSPSO-IS and P-MOEA/D-IS are depicted in the figures with (a), (b) and (c) respectively.

The EAF plot says that, in the epsilon dataset, P-NSGA-II-IS attained better diversity of solutions than the P-MOEA/D-IS, whereas the highest AUC is attained by the P-MOEA/D-IS. In the IEEE malware dataset, P-NSGA-II-IS stands top by maintaining diversity as well as the highest AUC. P-MOEA/D-IS stands second in this case. In both the epsilon and IEEE malware, P-NSPSO-IS stands last. In the case of the OVA_Omentum dataset, P-NSPSO-IS attained higher AUC but selected a higher number of features when compared with the P-NSGA-II-IS and P-MOEA/D-IS. P-NSGA-II-IS stands second in this case in terms of the cardinality of the feature subset but attained relatively less AUC when compared with P-NSPSO-IS. In the case of the OVA_Uterus dataset, P-NSGA-II-IS attained higher AUC with very less cardinal feature subset when compared with NSPSO. In both these OVA_Omentum and OVA_Uterus cases, P-MOEA/D-IS stands the last, both in terms of diversity and higher AUC. In the case of DDoS dataset case, both the P-NSPSO-IS and P-MOEA/D-IS achieved higher AUC than the P-NSGA-II-IS. It is important to note that P-NSPSO-IS achieved higher AUC by selecting a relatively lesser number of features than the P-MOEA/D-IS.

Overall, there is no clear winner. Yet P-NSGA-II-IS has achieved better performance by standing first or second except in the DDoS dataset. This shows that P-NSGA-II-IS has shown better exploration and exploitation capabilities than the latter two algorithms. Along with this, the EAF plots of P-NSGA-IS show that the former attained better diversity. P-NSPSO-IS performed well with the high dimensional datasets and DDoS. It failed miserably in the moderate dimensional datasets, i.e., Epsilon and IEEE malware, thereby getting stuck at the local optima. Note that even though the P-NSPSO-IS attained a better AUC than the other two algorithms, it also selected a very high number of features compared with other two algorithms. The EAF plots show that P-NSPSO-IS has attained lesser diversity than the other two algorithms. P-MOEA/D-IS failed miserably in the high dimensional dataset cases, thereby getting stuck at the local optima. P-MOEA/D-IS did well in the Epsilon dataset whereas achieved comparable AUC in the DDoS dataset but with more features than the P-NSPSO-IS. Overall, P-MOEA/D-IS has attained poor diversity.

## 6.3 Three-way comparative analysis

The 3-way comparative analysis is conducted between the three algorithms to understand their limitations, pros, and cons in the island based framework. The top solutions achieving the highest AUC obtained across the experiment runs are considered, and the mean AUC and mean cardinality is presented in Table 6.

P-MOEA/D-IS has achieved a slightly better mean AUC with a slightly more number of features than the P-NSGA-II-IS in the Epsilon dataset. In the IEEE Malware dataset, P-NSGA-II-IS did exceptionally well in AUC than the other two algorithms. In the DDoS dataset, P-NSGA-II-IS is outperformed by both the P-NSPSO-IS and P-MOEA/D-IS. P-NSPSO-IS achieved better AUC with a lesser number of features, and the P-MOEA/D-IS stands second in the list. In the OVA_Omentum, P-NSPSO-IS achieved a very high AUC than the latter two, but it was selected with a higher number of features than the other two algorithms. P-NSGA-II-IS achieved a better mean AUC in the OVA_Uterus dataset than the P-NSPSO-IS and P-MOEA/D-IS. The insights obtained from EAF are also replicated here. Hence, overall, P-NSGA-II-IS performed well in most cases by standing either first or second in the list.

Table 6: Mean cardinality and mean AUC obtained

| Dataset | P-NSGA-II-IS | | P-NSPSO-IS | | P-MOEA/D-IS | |
|---|---|---|---|---|---|---|
| | Mean cardinality | Mean AUC | Mean cardinality | Mean AUC | Mean cardinality | Mean AUC |
| Epsilon | 525.57 | 0.8101 | 237.78 | 0.5092 | **533.68** | **0.8192** |
| IEEE Malware | **335.52** | **0.8125** | 189.94 | 0.5332 | 224.42 | 0.7846 |
| OVA_Omentum | 25.27 | 0.7630 | **329.94** | **0.9668** | 81.07 | 0.5828 |
| OVA_Uterus | **68.15** | **0.8623** | 327.26 | 0.821 | 93.38 | 0.710 |
| DDoS | 4.55 | 0.414 | **7.94** | **0.987** | 19.4 | 0.9798 |

## 6.4 Repeatability

In order to evaluate the robustness and stability of the underlying wrapper, repeatability has occupied its importance. The more an optimal feature or feature subset repeats itself, the more powerful the underlying evolutionary algorithm is said to be. In this section, the repeatability is evaluated on the top solutions obtained over the 20 runs. The results are presented in Table 7.

In the case of the Epsilon dataset, P-NSGA-II-IS achieved 0.8091 with 521 features, whereas P-MOEA/D-IS stands top by achieving better 0.8194 with 513, which is a relatively lesser number of features than the former. Even though in the IEEE Malware dataset, P-MOEA/D-IS and P-NSPSO-IS selected lesser cardinal feature subset, but achieved lesser feature subset than the P-NSGA-II-IS, which has achieved 0.8091 with the feature subset having cardinality 257. In the OVA_Omentum, P-NSPSO-IS achieved 0.952, with 353 features standing first in the list. In the case of OVA_Uterus, P-NSGA-II-IS achieved 0.846 with 61 features. Even though in the case of the DDoS dataset, both the P-NSPSO-IS and P-MOEA/D-IS achieved almost similar AUC but the former achieved with relatively lesser number of features than the P-NSGA-II-IS.

Table 7: Cardinalities and the corresponding AUC of the Top-most repeated feature subsets

| Dataset | P-NSGA-II-IS | | P-NSPSO-IS | | P-MOEA/D-IS | |
|---|---|---|---|---|---|---|
| | #s1 | AUC | #s1 | AUC | #s1 | AUC |
| Epsilon | 521 | 0.8091 | 192 | 0.507 | **513** | **0.8194** |
| IEEE Malware | **257** | **0.8091** | 130 | 0.5 | 174 | 0.779 |
| OVA_Omentum | 32 | 0.7725 | **353** | **0.952** | 81 | 0.610 |
| OVA_Uterus | **67** | **0.844** | 288 | 0.818 | 86 | 0.705 |
| DDoS | 5 | 0.3266 | **5** | **0.9997** | 12 | 0.9901 |

* where #s1 is the cardinality of the top-most repeated feature subset

## 6.5 Least Cardinal Feature Subset with highest AUC

This subsection focuses primarily on the least cardinal feature subset with the highest AUC. Hence, here the algorithm which achieved highest AUC with the least cardinal feature subset is considered as the best algorithm. The results are presented in Table 8. The results prove that there is no clear winner, yet P-NSGA-II-IS achieved better in most of the cases except in the DDoS dataset. In the case of the DDoS dataset, P-NSPSO-IS achieved 0.9997 AUC with 5 (five) features, whereas the P-MOEA/D-IS achieved 0.9996 but with 8 (eight) features. In this case, P-NSGA-II-IS may have failed due to the lack of diversity in the population which would have contributed it spoor performance. In the case of Epsilon, P-MOEA/D-IS outperformed the other two algorithms by selecting a lower cardinal feature subset. In the case of IEEE, even though P-MOEA/D-IS and P-NSPSO-IS selected a lesser cardinal feature subset than the P-NSGA-II-IS, the latter has attained better AUC than the former two algorithms. In the case of OVA_Omentum and OVA_Uterus, P-NSPSO-IS has selected high cardinal feature subset than the other two algorithms. But it achieved higher AUC in OVA_Omentum but stood second in the OVA_Uterus dataset. In the case of the OVA_Uterus dataset, P-NSGA-II-IS performed better. In both these cases, P-MOEA/D-IS performed the least.

Table 8: Least Cardinal Feature subset selected by each approach with the most repetitions

| Dataset | P-NSGA-II-IS | | P-NSPSO-IS | | P-MOEA/D-IS | |
|---|---|---|---|---|---|---|
| | #s | AUC | #s | AUC | #s | AUC |
| Epsilon | 419 | 0.8095 | 161 | 0.5098 | **348** | **0.8201** |
| IEEE Malware | **257** | **0.8091** | 124 | 0.5 | 166 | 0.7799 |
| OVA_Omentum | 16 | 0.7430 | **280** | **0.9526** | 81 | 0.6108 |
| OVA_Uterus | **62** | **0.9080** | 288 | 0.8184 | 79 | 0.7245 |
| DDoS | 2 | 0.3266 | **5** | **0.9997** | 8 | 0.9996 |

* where #s is the cardinality of the feature subset having least cardinal subset with highest AUC

## 6.6 Speedup

Table 9: Speedup Analysis of parallel versions over the sequential ones

| Dataset | Algorithm | Sequential E.T | Parallel E.T | S.U |
|---|---|---|---|---|
| Epsilon | P-NSGA-II-IS | 20727.29 | 7307.64 | 2.83 |
| | P-NSPSO-IS | 16931.63 | 5491.11 | **3.08** |
| | P-MOEA/D-IS | 30897.56 | 10235.45 | 3.01 |
| IEEE Malware | P-NSGA-II-IS | 21975.40 | 7229.49 | **3.03** |
| | P-NSPSO-IS | 13789.33 | 5703.92 | 2.41 |
| | P-MOEA/D-IS | 21290.891 | 7244.14 | 2.93 |
| OVA_Omentum | P-NSGA-II-IS | 5630.99 | 1761.18 | 3.19 |
| | P-NSPSO-IS | 13758.28 | 4796.52 | 2.86 |
| | P-MOEA/D-IS | 7197.8 | 2201.73 | **3.26** |
| OVA_Uterus | P-NSGA-II-IS | 5269.18 | 1965.93 | 2.68 |
| | P-NSPSO-IS | 13538.45 | 4581.132 | 2.95 |
| | P-MOEA/D-IS | 6248.56 | 2095.87 | **2.98** |
| DDoS | P-NSGA-II-IS | 14781.49 | 5349.56 | 2.76 |
| | P-NSPSO-IS | 14414.62 | 5124.97 | 2.81 |
| | P-MOEA/D-IS | 14134.71 | 4528.98 | **3.12** |

* where E.T is the execution time given in seconds, S.U is the speedup achieved.

The ratio between the time taken by the sequential version and the time taken by the parallel version is defined as the speedup. The mathematical representation is given in Eq. (9). This has occupied its importance in determining the performance of the parallel version of the algorithm. The results are presented in Table 9.

$$\text{Speed Up (S.U)} = \frac{\text{Time taken by Sequential Version}}{\text{Time taken by Parallel Version}} \quad (9)$$

Speedup stands as one of the essential characteristics in evaluating the performance of the parallel version of the algorithm. The results are presented in Table 9. We observed that the proposed parallel algorithms achieved significant speedup. Overall the speedup achieved ranges from 2.41 – 3.26 times by all proposed algorithms over their sequential counterparts in all datasets. As the number of nodes in the cluster is 5, the maximum possible speedup that could be achieved is 5. The linear speedup is not achieved because the underlying parallel model is designed with synchronization junctions (in-migration).

## 6.7 Diversity preservation using Hypervolume

The 3-way comparative analysis is conducted to measure the state of convergence and extent of diversity found by bi-objective evolutionary algorithms. We analysed the diversity preservation using Hypervolume (HV) metric.

Hypervolume calculates volume of the dominated set of solutions, w.r.t a reference point. Here the reference point is the worst possible solution. It helps in understanding how well the diversity is preserved. The algorithm having higher hypervolume tends to have better diversity.

HV is calculated for each of the algorithms across all the datasets as follows: For each run, HV is calculated on the obtained evolved population. As mentioned earlier, the number of experiment runs arrived to be 20. Hence, for each run, we computed one HV. The mean HV value results are presented in Table 10. It is observed that P-NSPSO-II-IS has shown better diversity than the rest of the algorithms in the case of Epsilon and IEEE Malware datasets. However, P-NSGA-II-IS has shown better diversity in the high dimensional datasets such as OVA_Omentum and OVA_Uterus datasets and DDoS dataset. Overall, P-MOEA/D-IS has shown poor diversity in almost all cases.

Table 10: Mean HV obtained by all algorithms over 20 runs

| Dataset | Mean HV of P-NSGA-II-IS | Mean HV of P-NSPSO-II-IS | Mean HV of P-MOEA/D-IS |
|---|---|---|---|
| Epsilon | 0.379 | **0.513** | 0.202 |
| IEEE Malware | 0.407 | **0.527** | 0.216 |
| OVA_Omentum | **0.508** | 0.282 | 0.474 |
| OVA_Uterus | **0.590** | 0.374 | 0.377 |
| DDoS | **0.804** | 0.351 | 0.351 |

## 6.7 Statistical testing of the results

This section analyzes the statistical testing done w.r.t the Hypervolume (HV) as this metric is suitable for the bi-objective EAs.

The pairwise t-test is conducted based on the HV scores achieved by each of the algorithms in the 20 experiment runs. The following pairwise t-test analysis emphasizes the statistically valid statements about the performance across various datasets. The results are presented in Table 9. The significance level is chosen as 5%, and the degrees of freedom is taken as 38 (20+20-2).

The null hypothesis and the alternate hypothesis are taken as follows:

*H₀: both the algorithms are statistically equal*
*H₁: both the algorithms are statistically not equal.*

As the p-values for all datasets are less than 0.05, the null hypothesis is rejected, and the alternate hypothesis is accepted. From the t-statistic and the p-value presented in Table 11, we infer that all the three algorithms are significantly performed differently from each other.

In most of the cases, P-NSGA-II-IS performed significantly better than the P-NSPSO-IS and also P-MOEA/D-IS. Especially in the case of DDoS, P-NSGA-II-IS doesn't perform significantly than the P-NSPSO-IS and P-MOEA/D-IS. In the case of OVA_Omentum, P-NSGA-II-IS performed significantly better than the P-MOEA/D-IS but failed to do with the P-NSPSO-IS.

It is important to note that P-NSPSO-IS and P-MOEA/D-IS behaved similarly in the DDoS dataset. In the rest of the datasets, P-MOEA/D-IS is significant in the Epsilon and IEEE Malware datasets, whereas, in the case of very high dimensional datasets, P-NSPSO-IS is statistically significant better than the P-MOEA/D-IS.

Table 11: Pairwise T-test analysis: A three way comparison based on HV

| Dataset | P-NSGA-II-IS vs P-NSPSO-IS | | | P-NSGA-II-IS vs P-MOEA/D-IS | | | P-NSPSO-IS vs P-MOEA/D-IS | | |
|---|---|---|---|---|---|---|---|---|---|
| | t-statistic | p-value | T/F | t-statistic | p-value | T/F | t-statistic | p-value | T/F |
| Epsilon | 258.53 | $1.90 \times 10^{-60}$ | T | 7.40 | $9.63 \times 10^{-9}$ | T | 330.58 | $2.73 \times 10^{-64}$ | T |
| IEEE Malware | 28.765 | $1.95 \times 10^{-26}$ | T | 18.03 | $1.3 \times 10^{-19}$ | T | 25.69 | $9.43 \times 10^{-25}$ | T |
| OVA_Omentum | 24.49 | $1.28 \times 10^{-23}$ | T | 16.75 | $1.87 \times 10^{-16}$ | T | 54.79 | $1.23 \times 10^{-31}$ | T |
| OVA_Uterus | 3.23 | 0.002 | T | 9.957 | $6.95 \times 10^{-12}$ | T | 12.38 | $1.53 \times 10^{-14}$ | T |
| DDoS | 40.009 | $2.06 \times 10^{-43}$ | T | 40.615 | $2.66 \times 10^{-44}$ | T | 1.403 | 0.168 | **F** |

# 7. Conclusions and limitations

In a first-of-its-kind study, we developed and proposed island based MOEAs with a migration strategy under the spark environment. This study developed the parallel versions of NSGA-II, MOEA/D, and NSPSO based wrappers for the FSS. Throughout, LR is chosen as the classifier. We demonstrated the effectiveness of MapReduce on five high-dimensional datasets. The results indicate that there is no clear winner, yet P-NSGA-II-IS has performed better in most cases. P-NSPSO-IS also performed well, and P-MOEA/D-IS performed the least. The statistical analysis says that the P-NSGA-II-IS is significantly different compared to P-NSPSO-IS and P-MOEA/D-IS at both exploration and exploration.

The limitations of the current work are as follows: (i) The current study is conducted in a bi-objective function environment where cardinality and AUC are considered as two objective functions. (ii) The parallel EA based wrappers are non-adaptive. (iii) The other popular variants of various MOEAs can be tried out in the framework, and especially trying indicator-based MOEAs stands as the potential work to be carried out in the future. (iv) The current parallel strategy is designed with a synchronization junction in a migration rule. As regards future work, any algorithm from the Table 1 or any hybrid combination of a few algorithms can also be designed under spark environment. Designing asynchronous parallel methods (or) adopting generation parallelizing [13] is another potential future work. Further, investigation will be carried out on the same problem set but in many-objective environment. Moreover, in immediate future, we plan to develop newer variants of the parallel MOEAs presented here by incorporating the concepts of chaos and quantum computing in them.

# Appendix

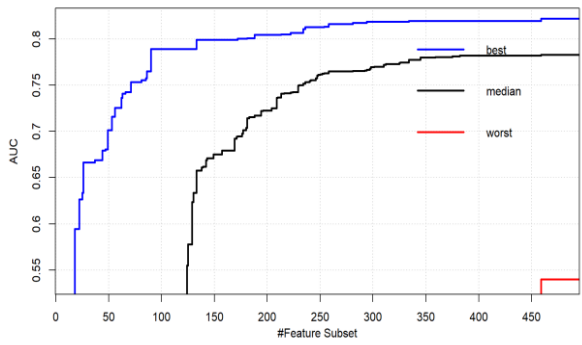
Fig. A.1(a) P-NSGA-II-IS EAF on the Epsilon dataset

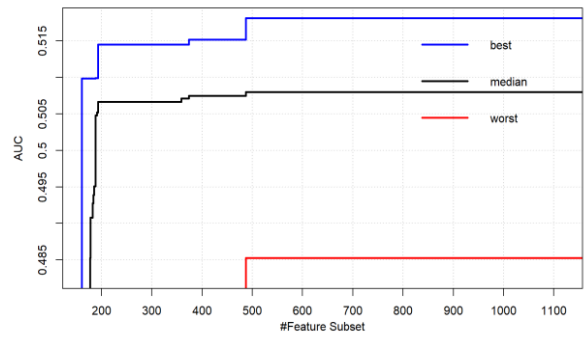
Fig. A.1(b) P-NSPSO-IS EAF on the Epsilon dataset

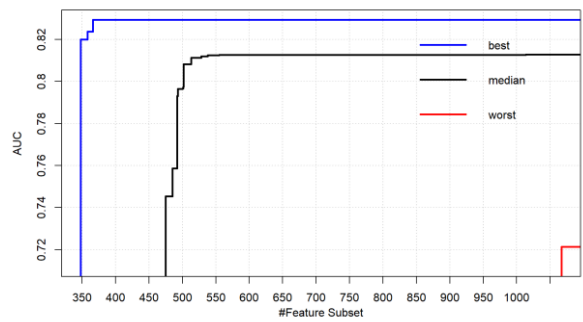
Fig. A.1 (c) P-MOEA/D-IS EAF on the Epsilon dataset

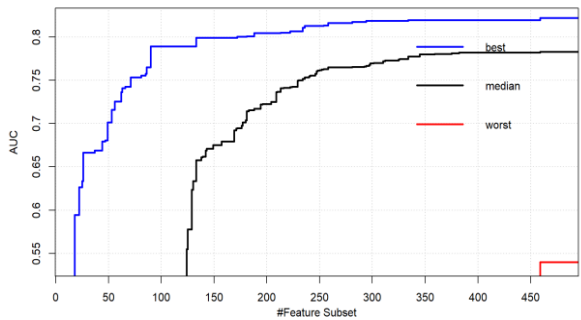
Fig. A.2(a) P-NSGA-II-IS EAF on the IEEE Malware dataset

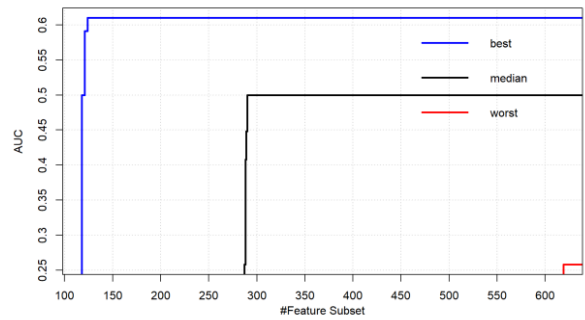
Fig. A.2(b) P-NSPSO-IS EAF on the IEEE Malware dataset

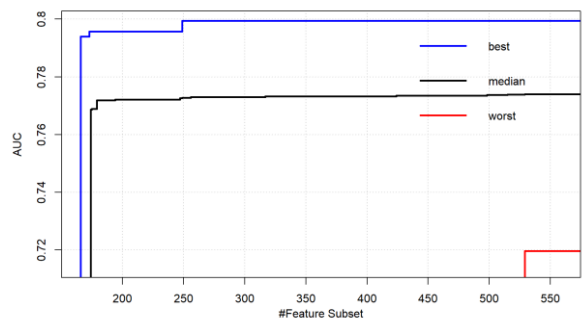
Fig. A.2(c) P-MOEA/D-IS EAF on the IEEE Malware dataset

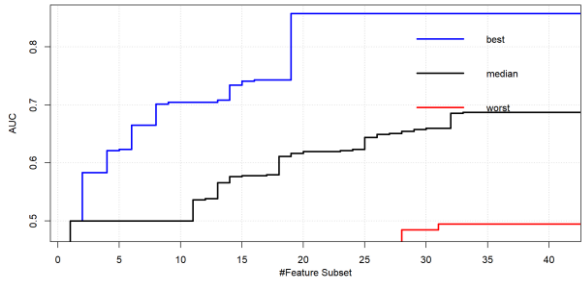

Fig. A.3(a) P-NSGA-II-IS EAF on the OVA_Omentum dataset

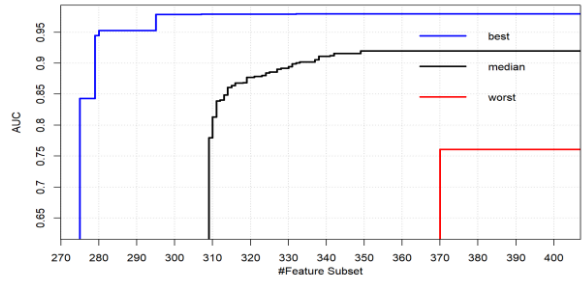

Fig. A.3(b) P-NSPSO-IS EAF on the OVA_Omentum dataset

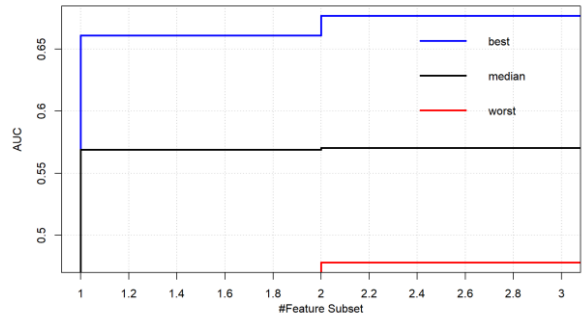

Fig. B.3(c) P-MOEA/D-IS EAF on the OVA_Omentum dataset

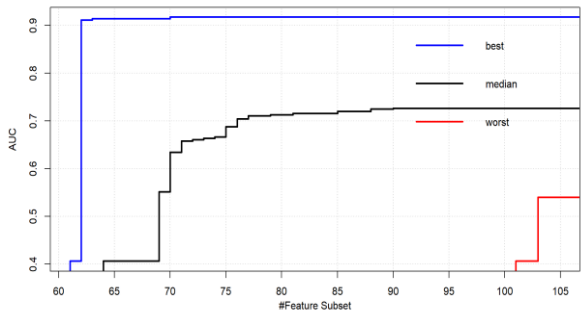

Fig. A.4(a) P-NSGA-II-IS EAF on the OVA_Uterus dataset

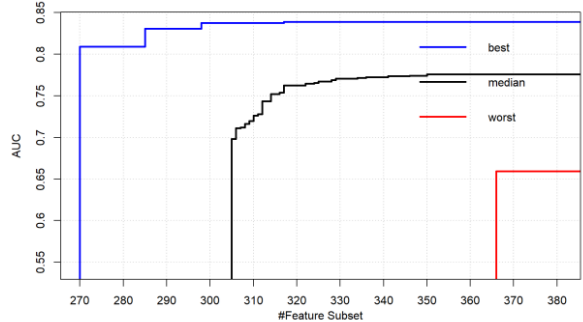

Fig. A.4(b) P-NSPSO-IS EAF on the OVA_Uterus dataset

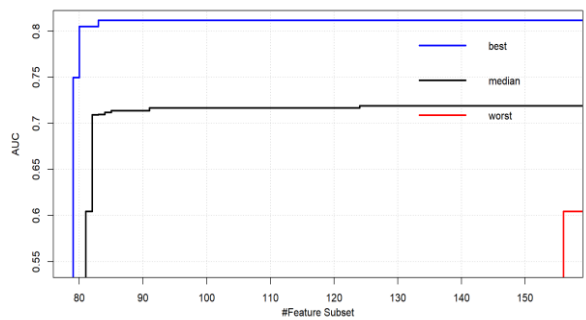

Fig. A.4(c) P-MOEA/D-IS EAF on the OVA_Uterus dataset

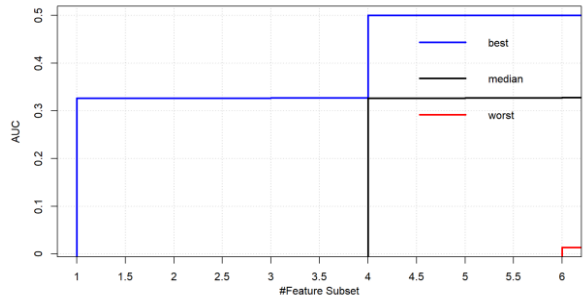 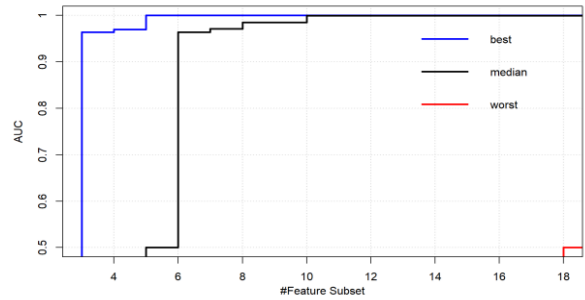

Fig. A.5(a) P-NSGA-II-IS EAF on the DDoS dataset

Fig. A.5(b) P-NSPSO-IS EAF on the DDoS dataset

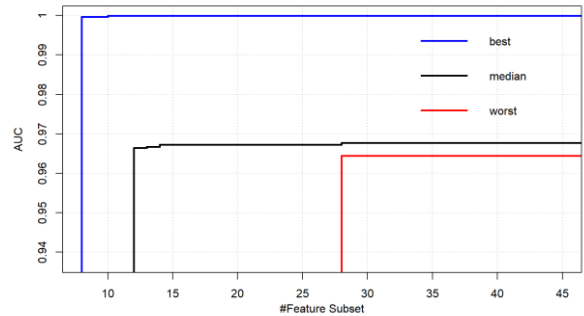

Fig. A.5(c) P-MOEA/D-IS EAF on the DDoS dataset